\documentclass[twocolumn]{svjour3} 
\usepackage{amsmath}  					
\usepackage{amssymb}  					
\usepackage{xcolor}
\usepackage{graphicx}
\usepackage{soul}
\usepackage{cite}
\usepackage{doi}
\usepackage{hyperref}

\newcommand{\bs}[1]{\boldsymbol{#1}}

\newcommand{\half}{\frac{1}{2}}

\newcommand{\be}{\begin{equation}}
\newcommand{\ee}{\end{equation}}

\newcommand{\mth}{$m^{\text{th}}\,$}

\newcommand{\dn}{\text{dn}}
\newcommand{\sn}{\text{sn}}
\newcommand{\cn}{\text{cn}}

\RequirePackage{fix-cm}
\usepackage{subfig}

\journalname{Nonlinear Dynamics}
\begin{document}
	
	\title{Higher-order Breathers as Quasi-rogue Waves on a Periodic Background}
	
	\author{Omar A. Ashour \and Siu A. Chin %
	        \and Stanko N. Nikoli\'c \and %
	        Milivoj R. Beli\'c}
	
	\institute{O. Ashour \at
              Department of Physics, University of California, Berkeley CA 94720, USA \\
              \email{ashour@berkeley.edu}           
           \and
           S. Chin \at
                Department of Physics and Astronomy, Texas A\&M University, College Station, TX 77843, USA
           \and
           S. Nikoli\'c \at
           Institute of Physics, University of Belgrade, Pregrevica 118, 11080 Belgrade, Serbia \\
            Science Program, Texas A\&M University at Qatar, P.O. Box 23874 Doha, Qatar
           \and
           M. Beli\'c \at
            Science Program, Texas A\&M University at Qatar, P.O. Box 23874 Doha, Qatar
    }

    \date{Received: date / Accepted: date}
	
	\maketitle
	
	\begin{abstract}    
		We investigate higher-order breathers of the cubic nonlinear Schr\"odinger equation on an elliptic background. We find that, beyond first-order, any arbitrarily constructed breather is a single-peaked solitary wave on a disordered background. These ``quasi-rogue waves" are also common on periodic backgrounds. We assume the higher-order breather is constructed out of constituent first-order breathers with commensurate periods (i.e., higher-order harmonic waves). In that case, one obtains ``quasi-periodic" breathers with distorted side-peaks. Fully periodic breathers are obtained when their wavenumbers are harmonic multiples of the background and each other. They are truly rare, requiring finely-tuned parameters. Thus, on a periodic background, we arrive at the paradoxical conclusion that the apparent higher-order rogue waves are rather common, while the truly periodic breathers are exceedingly rare.
		

	\end{abstract}
	
\section{Introduction \label{sec:intro}}

The dimensionless cubic NLS equation, given by
\begin{equation}
i \frac{\partial \psi}{\partial x} + \half \frac{\partial^2 \psi}{\partial t^2} + \psi\left|\psi\right|^2 = 0,
\label{eq:nlse}
\end{equation}
with $t$ and $x$ being the transverse and longitudinal variables, and $\psi \equiv \psi(x,t)$ the slowly-varying wave envelope, has been widely used in the field of nonlinear optics and photonics to guide experimental realizations \cite{Solli2007,Frisquet2013,Armarolia} and theoretical explorations \cite{Onorato2013,Akhmediev2009,Kedziora2012a,Kedziora2011,Kedziora2014,Kedziora2012,Akhmediev2009a} of higher order breathers and rogue waves in optical fibers. For a recent review, see Ref. \cite{Dudley2014}.

While the study of breathers and rogue waves
on a uniform background, including those based on extended NLS equations, has matured with a body of analytical results \cite{Chin2015,Akhmediev2009,Dudley2014,Kedziora2011,Chin2016, Ankiewicz2011a,Chowdury2015a,Ankiewicz2016,Kedziora2015}, similar studies on periodic backgrounds have started only recently \cite{Kedziora2014,Chin2016a,Nikolic2017, Ashour2017}. 

In this work, we are specifically interested in breathers on the dnoidal Jacobi elliptic function (JEF) background, which has the form:
\begin{align}
\psi_0(x,t) = \dn(t; k)e^{i x(1-\frac{k^2}{2})}, \label{eq:dnseed}
\end{align}
where $k$ is the elliptic \emph{modulus} (not to be confused with the elliptic \emph{parameter}, $k^2$). As our analysis will show below, at this time it is unlikely that one can derive analytical expressions for our breathers; therefore we focus on studying solutions of \eqref{eq:nlse} using a numerical implementation of the analytical Darboux transformation (DT) procedure \cite{Kedziora2012a,Kedziora2011,Kedziora2014,Kedziora2012,Akhmediev2009a}. In a nutshell, the procedure states that given a simple, zeroth-order ``seed" solution $\psi_0(x,t)$ of \eqref{eq:nlse}, such as \eqref{eq:dnseed}, one can generate an $N^\text{th}$-order solution recursively via:
\begin{align}
\psi_N(x,t) = \psi_0(x,t) + \sum_{m=1}^{N}\frac{2 i r_{m1} s_{m1}^* \left(\lambda_m -\lambda_m ^*\right)}{|r_{m1}|^2+|s_{m1}|^2},
\label{eq:DTIter}
\end{align}
where the sum goes over $N$ constituent first-order solutions of the Lax pair equation, characterized by generally complex eigenvalues $\lambda_m$. The two functions $r_{m1}(x,t)$ and $s_{m1}(x,t)$ are the Lax pair generating functions for the $m^{\text{th}}$ first-order constituent breather. In this work, we restrict ourselves to purely imaginary eigenvalues:
\begin{align}
\lambda_m = i \nu_m, \quad \text{with} \quad 0 < \nu_m <1,
\end{align}
corresponding to Akhmediev type breathers \cite{Akhmediev1987b}, rather than Kuznetsov-Ma type breathers \cite{Kuznetsov1977,Ma1979}.
As per the Peak Height Formula (PHF) \cite{Chin2016a}, these solutions have a peak-height of:

\begin{equation}
\psi_N(0,0) = 1 +2 \sum_{m=1}^{N} \nu_m.
\label{eq:phfDn}
\end{equation}

In this work, we will concentrate on demonstrating the surprising finding mentioned in the abstract: beyond first-order, any arbitrarily constructed breather on an elliptic background looks very much like a rogue wave. We relegate all the technical details on the DT and the PHF to the appendix.

The paper is structured as follows. In Sec. \ref{sec:tranPeriodic} we demonstrate, by enforcing the reality of the wavenumber, that in contrast to the uniform background case \cite{Chin2015, Chin2016}, the first-order breathers on an elliptic background can at most be quasi-periodic in the $t$-direction unless the matching condition \eqref{eq:gammaContours} is obeyed. 
The concatenation of $N$ such first-order breathers then generally results in an  \emph{aperiodic} $N^\text{th}$-order breather greatly resembling a rogue wave (RW).  In Sec. \ref{sec:quasiPeriodic}, we show that if the periods of the constituent first-order breathers are commensurate with one another, then one sees a quasi-periodic $N^\text{th}$-order breather, where the repeating side peaks are distorted by the background. 
In Sec. \ref{sec:fullyPeriodic}, we show that truly periodic $N^\text{th}$-order breathers are produced if and only if the periods of the constituent first-order breathers are all matched to that of the background. This requires restricting the parameters $\nu$ and $ k$ to a special set of contours $\gamma_q$ (Eq. \eqref{eq:gammaContours}) in the $\nu k$-plane.  Our conclusions are summarized in Sec. \ref{sec:conclusion}.

\section{Aperiodic Breathers on an Elliptic Background \label{sec:tranPeriodic}}

For a breather to be periodic (along the $t$ direction), its wavenumber must be real \cite{Kedziora2014}. Assuming purely imaginary eigenvalues $\lambda_m = i\nu_m$, the half-wavenumber is given by:
\begin{align}
\kappa_m  =\sqrt{\left(\lambda_m-\frac{k^2}{4 \lambda_m}\right)^2+1}= 
\sqrt{1-\left(\nu_m+\frac{k^2}{4 \nu_m}\right)^2}.
\label{eq:kappa_dn}
\end{align}
For $\kappa_m$ to be real, we must have
\begin{equation}
(\nu_m+\frac{k^2}{4\nu_m})^2\le 1\quad\implies\quad k^2\le 4\nu_m(1-\nu_m),
\label{keq}
\end{equation}
which restricts the range of $\nu_m$ to
\begin{equation}
\frac12-\frac12\sqrt{1-k^2}\le\nu_m\le \frac12+\frac12\sqrt{1-k^2}.
\label{vran}
\end{equation}
Alternatively, from (\ref{keq}) one sees that
\begin{align}
k^2\le 4\Big[\frac12+(\nu_m-\frac12)\Big]\Big[\frac12-(\nu_m-\frac12)\Big],
\end{align}
or
\begin{align}
k^2+\frac{(\nu_m - \frac{1}{2})^2}{(1/2)^2} \le 1,
\end{align}
which then restricts $(\nu_m, k)$ to half of an ellipse $\Gamma$, centered
on (1/2,0), with the vertical semi-major axis being 1 and the horizontal
semi-minor axis 1/2. This is shown in Fig. \ref{fig:firstPeriodicity} below, and 
summarized succinctly by:
\begin{equation}
\begin{aligned}
(\nu_m, k) \in \Gamma := \left\{(\nu_m, k) \in (0,1)\times[0,1] \, \Bigg| k^2+\frac{(\nu_m - \frac{1}{2})^2}{(1/2)^2} \le 1 \right\}. \label{eq:fundPeriod}
\end{aligned}
\end{equation}

\begin{figure*}[h] 
	\centering
	\includegraphics[width=.95\linewidth]{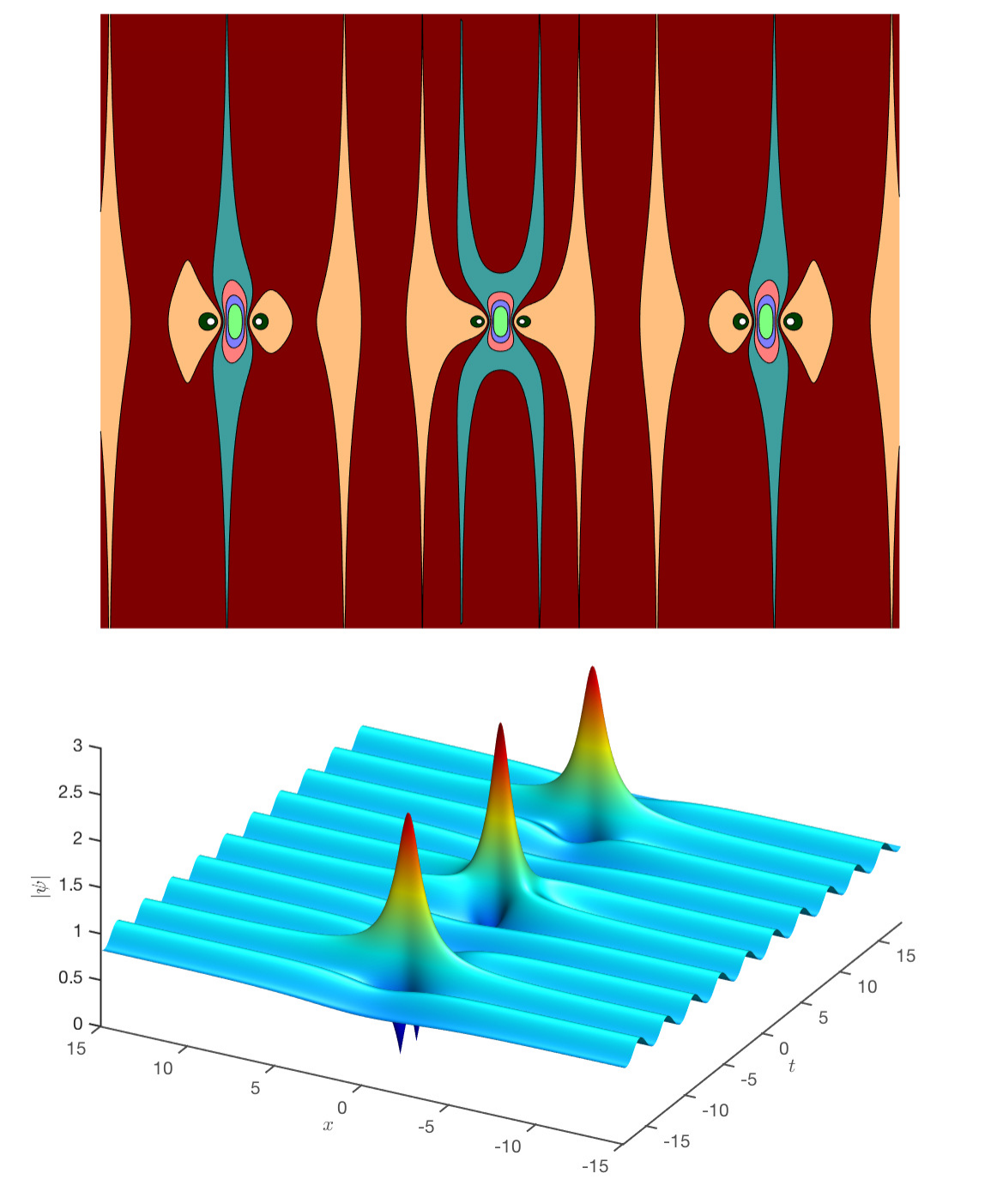}
	\caption{First-order breather with $(\nu = 0.8602, k = 3/5) \in \Gamma$ (i.e. $\kappa \in \mathbb{R}$). The distortion of the side peaks is evident in the contour plot. In all contour plots that follow, maroon represents the background ($|\psi|$ = 1), teal green is 0.25 \emph{above} the background, while light orange is 0.2 \emph{below} the background. The remaining colors denote other relevant low-$|\psi|$ features. The colormap is designed to showcase only the relevant low-$|\psi|$ features of the breathers, with a maximum height of 0.8 above the background. \label{fig:AB1_unmatched}}
\end{figure*}

\begin{figure*}[h] 
	\begin{minipage}{.5\linewidth}
		\centering
		\subfloat[]{\includegraphics[width=.95\linewidth]{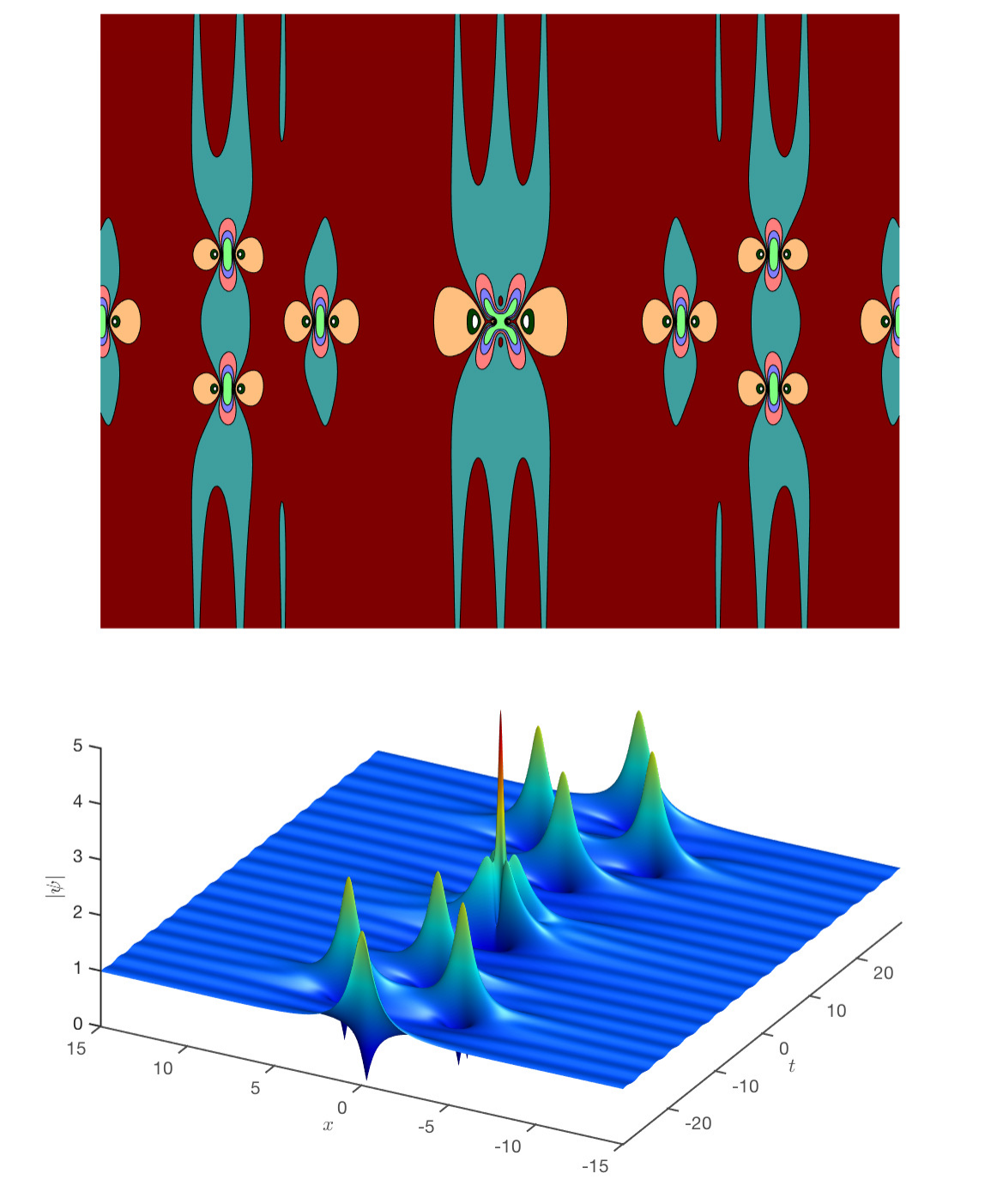}}
	\end{minipage}%
	\begin{minipage}{.5\linewidth}
		\centering
		\subfloat[]{\includegraphics[width=.95\linewidth]{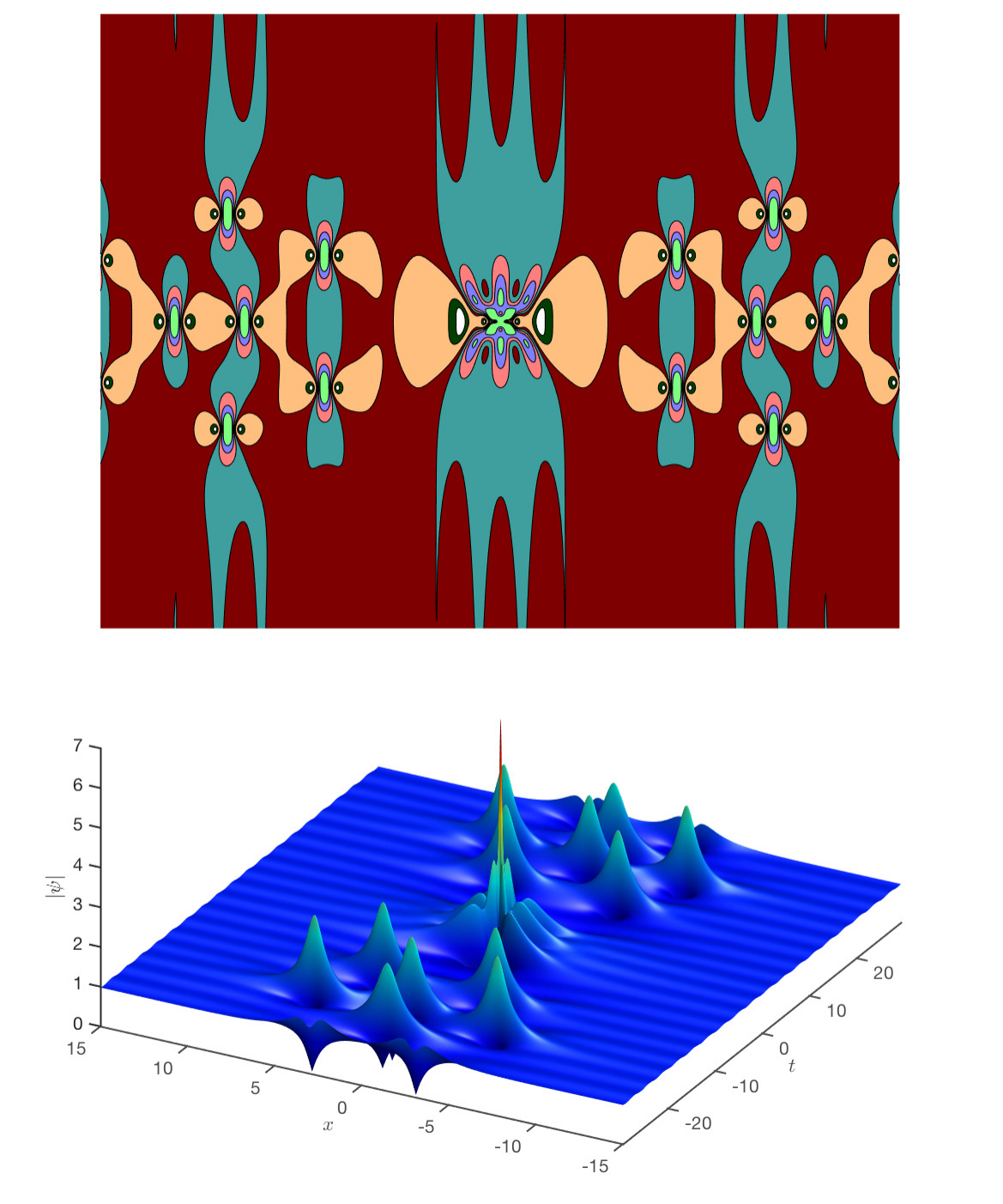}}
	\end{minipage}\par
	\begin{minipage}{.5\linewidth}
		\centering
		\subfloat[]{\includegraphics[width=.95\linewidth]{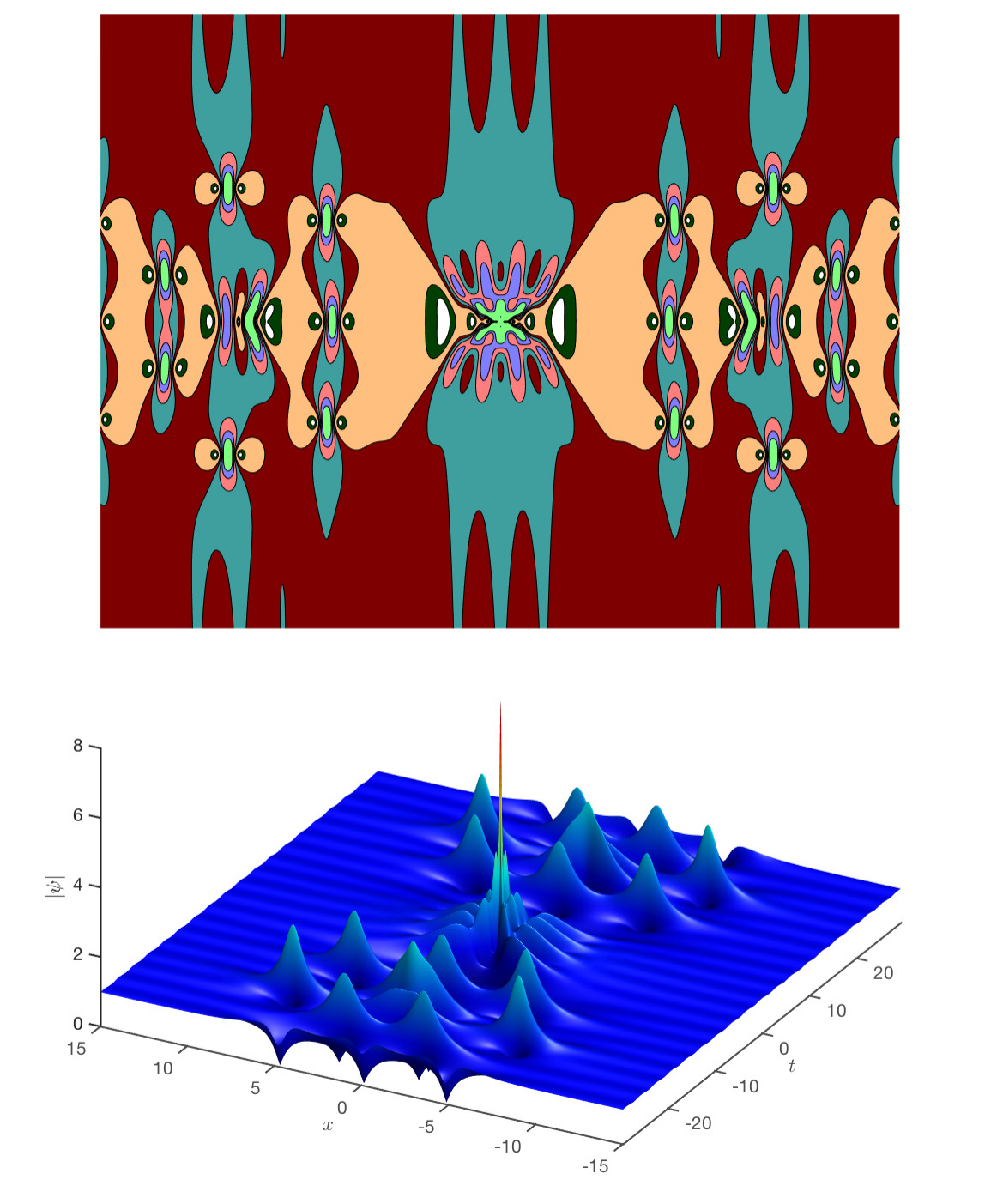}}
	\end{minipage}%
	\begin{minipage}{.5\linewidth}
		\centering
		\subfloat[]{\includegraphics[width=.95\linewidth]{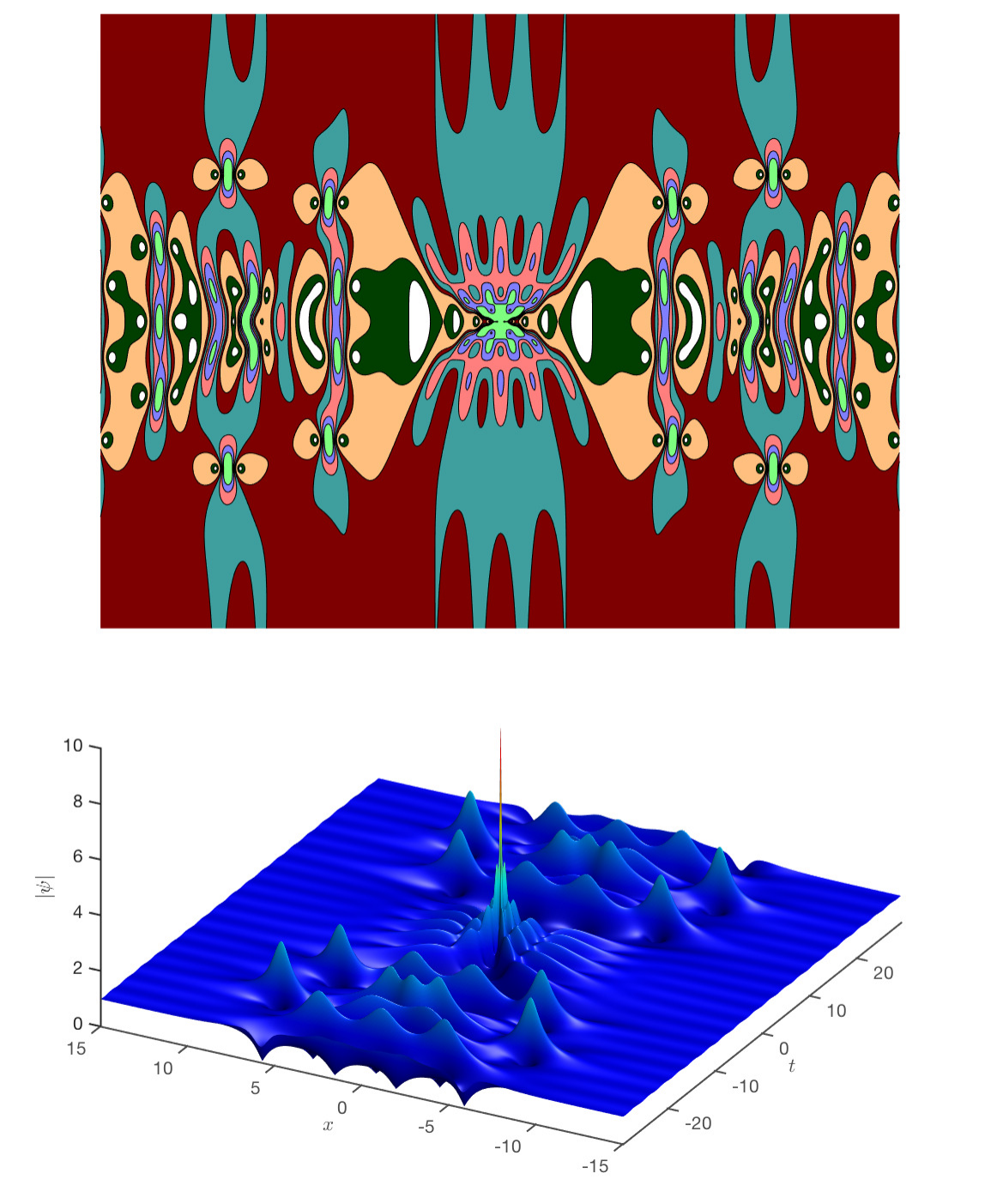}}
	\end{minipage}\par\medskip
	\caption{Higher-order breathers on a $\dn$ background with $k = 1/4$ and $\kappa_m \in \mathbb{R}$. (a) Second-order, with $\{\nu_m\} = \{0.98, 0.91\}$. (b) Third-order, with $\{\nu_m\} = \{0.98, 0.91, 0.84\}$ (c) Fourth-order, with $\{\nu_m\} = \{0.98, 0.91, 0.84,0.72\}$ (d) Fifth-order, with $\{\nu_m\} = \{0.98, 0.91, 0.84,0.72,0.51\}$. Inserts: Contour plots to emphasize the low-$|\psi|$ details. \label{fig:ABN_unmatched}}
\end{figure*}

In Fig. \ref{fig:AB1_unmatched}, we show the three periods of a first-order breather obeying the $\kappa$-reality condition \eqref{eq:fundPeriod}. This breather is not periodic since the repeating side-peaks are {\it not} identical to the central peak. Next, in Fig. \ref{fig:ABN_unmatched}, we show a set of higher-order breathers. Their eigenvalues $\nu_m$ obey \eqref{eq:fundPeriod}, but are otherwise arbitrary. These higher-order breathers are even more aperiodic, with the repeating side peaks highly distorted and diminished in intensity, leaving only a central peak of great height intact. It is a matter of definition how one defines a rogue wave, but Fig. \ref{fig:ABN_unmatched} clearly shows four solitary, high-intensity peaks amid an increasingly disordered background, each resembling more an oceanic rogue wave than a breather. We will refer to these unmatched solutions as {\it quasi-rogue waves} (QRWs). The phenomenon of QRW has only come to light recently in the study of higher-order breathers on periodic backgrounds \cite{Ashour2017, Nikolic2018}. They are not apparent in the study of first-order breathers on a periodic background, nor are they mentioned in the study of RWs on a constant background.

\section{Quasi-periodic Breathers \label{sec:quasiPeriodic}}

We now match the periods of the constituent breathers of these higher-order structures to the fundamental breather, i.e.:
\begin{align}
\kappa_m = m \kappa, \qquad m = 2, 3, ... N,
\label{eq:matchingBreathers}
\end{align}
where $\kappa \equiv \kappa_1$ is the period of the fundamental breather, characterized by $\nu \equiv \nu_1$. We choose the constituent breather wave numbers to be harmonics of each other, analogously to the uniform background case where these breathers would have been strictly periodic and of maximal peak-height (at a given periodicity) \cite{Chin2016}. Writing out \eqref{eq:matchingBreathers} explicitly, gives:
\begin{align}
m \sqrt{-\frac{k^4}{16 \nu ^2}-\frac{k^2}{2}-\nu ^2+1}=\sqrt{-\frac{k^4}{16 \nu _m^2}-\frac{k^2}{2}-\nu _m^2+1},
\end{align}
which leads to an expression for $\nu_m$ in terms of $\nu$ and $k$
\begin{align}
\nu_m(k, \nu) = \frac{\sqrt{G_m(k,\nu) + \sqrt{\left[G_m(k,\nu)\right]^2-64 k^4 \nu ^4}}}{4 \sqrt{2} \nu }\, , \label{eq:nu_m}
\end{align}
where
\begin{align}
G_m(k,\nu) \equiv k^4 m^2+8 \left(k^2-2\right) \left(m^2-1\right) \nu ^2+16 m^2 \nu ^4\,.
\end{align}

\begin{figure*}[h]
	\centering
	\includegraphics[width=0.9\linewidth]{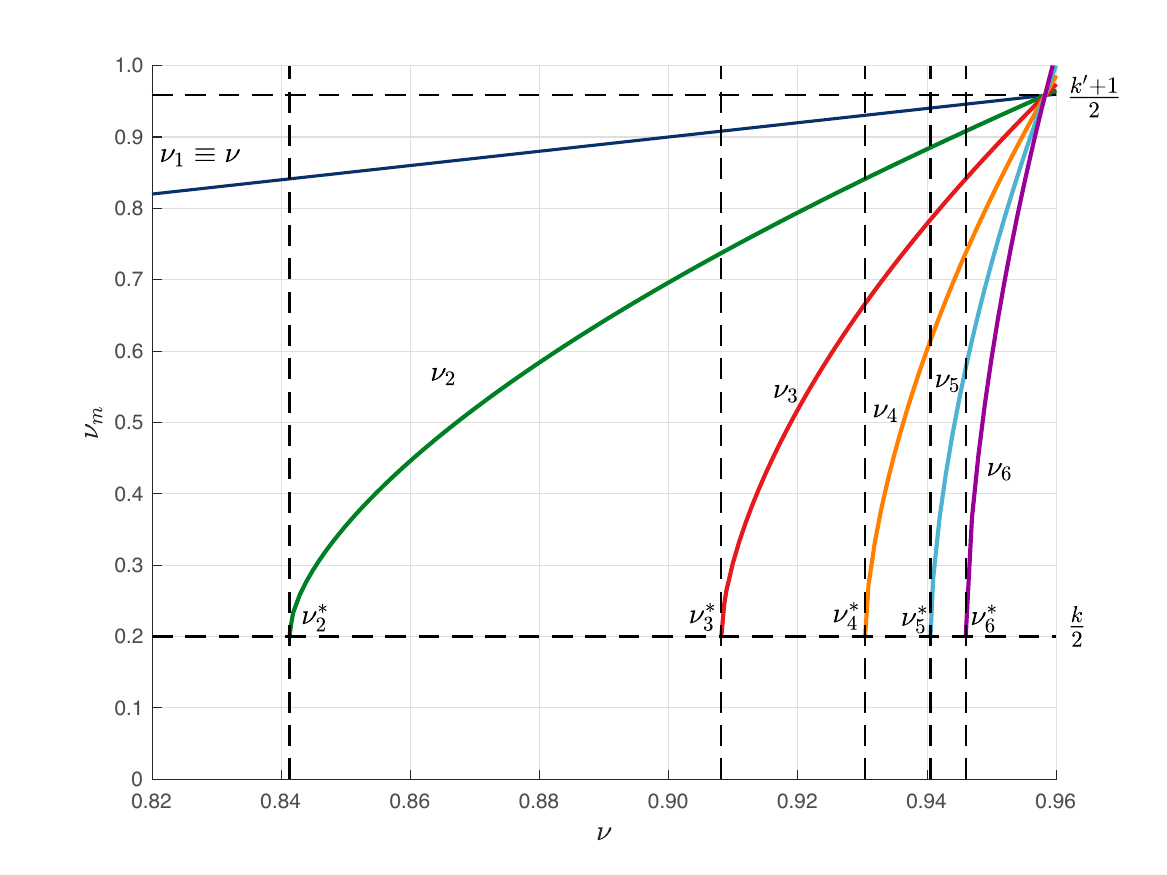}
	\caption{Plot of the eigenvalues $\nu_m$ given by \eqref{eq:nu_m} as a function of $\nu$, with $k=2/5$. The dashed vertical lines represent the lower limit $\nu_m^*$ (Eq. \ref{eq:nudn_star}), and the horizontal lines give the upper and lower limit on $\nu_m$. \label{fig:uglyDnFamily}}
\end{figure*}

A plot of Eq. \eqref{eq:nu_m} for multiple values of $m$ is shown in Fig. \ref{fig:uglyDnFamily}. Note that in Fig. \ref{fig:uglyDnFamily}, the curves $\nu_m(k)$, $\forall~m~\in~\mathbb{Z}^{+},$ clearly intersect at a point $(\nu_{\text{max}}, \nu_\text{max})$ in the $\nu\nu_m$-plane, given by the upper limit of Eq. \eqref{vran}:
\begin{align}
\nu_{\text{max}}(k) = \frac{k'}{2}+\frac{1}{2}\,, \label{eq:nu_max}
\end{align}
where $k' \equiv \sqrt{1-k^2}$ is the complementary elliptic modulus. A first-order solution characterized by this value of $\nu$ is known as the Concentrated Cnoidal Rogue Wave (CCRW) \cite{Kedziora2014} and was simply referred to as the``bright" rogue wave in Ref. \cite{Chin2016a}. This solution with $\kappa = 0$ is a single solitary peak on a periodic background, and is considered the natural generalization of the Peregrine rogue wave on a constant background. However, higher-order rogue waves generated by multiple values of \eqref{eq:nu_max} have degeneracy problems and cannot be described by the DT \cite{Kedziora2012}. By contrast, as shown in Fig. \ref{fig:ABN_unmatched}, an arbitrary high order QRW can be easily constructed using DT.

Figure \ref{fig:uglyDnFamily}, showing the plot of Eq. \eqref{eq:nu_m}, implies that for a fixed $m$, there is some cutoff $\nu = \nu_m^*(k)$ such that $\nu_l(\nu = \nu_m^*, k) \notin \mathbb{R} \, \forall \, l > m$. Thus, to generate a breather of order at most $N$, one must have $\nu \in (\nu_N^*, \nu_{N+1}^*]$. The imaginary part of $\nu_m(k, \nu)$ will be non-zero if and only if the term under the inner square root in \eqref{eq:nu_m} is negative. Thus, to find $\nu_m^*$, we need to solve:
\begin{align}
\left[G_m(k,\nu = \nu_m^*)\right]^2-64 k^4 (\nu_m^*)^4 = 0\,.
\end{align}
This gives:
\begin{align}
\nu_m^*(k) = \frac{\sqrt{2 C_m(k)+H_m(k)}}{2m} \, ,
\label{eq:nudn_star}
\end{align}
where
\begin{align}
\begin{aligned}
C_m(k)&=\sqrt{\left(m^2-1\right) \left(k'\right)^2 \left(m^2-\left(k'\right)^2\right)}\,, \\ 
H_m(k)&=m^2 \left(\left(k'\right)^2+1\right)-2 \left(k'\right)^2\,.
\end{aligned}
\end{align}
Substituting \eqref{eq:nudn_star} into \eqref{eq:nu_m}, one obtains $\nu_m(\nu = \nu_m^*, k) = k/2,\,\,\forall\, m \, \in \, \mathbb{Z}^{+}$, independent of $m$. The importance of this result will be discussed in Appendix \ref{sec:phf}.

\begin{figure*}[h]
	\centering
	\includegraphics[width=0.9\linewidth]{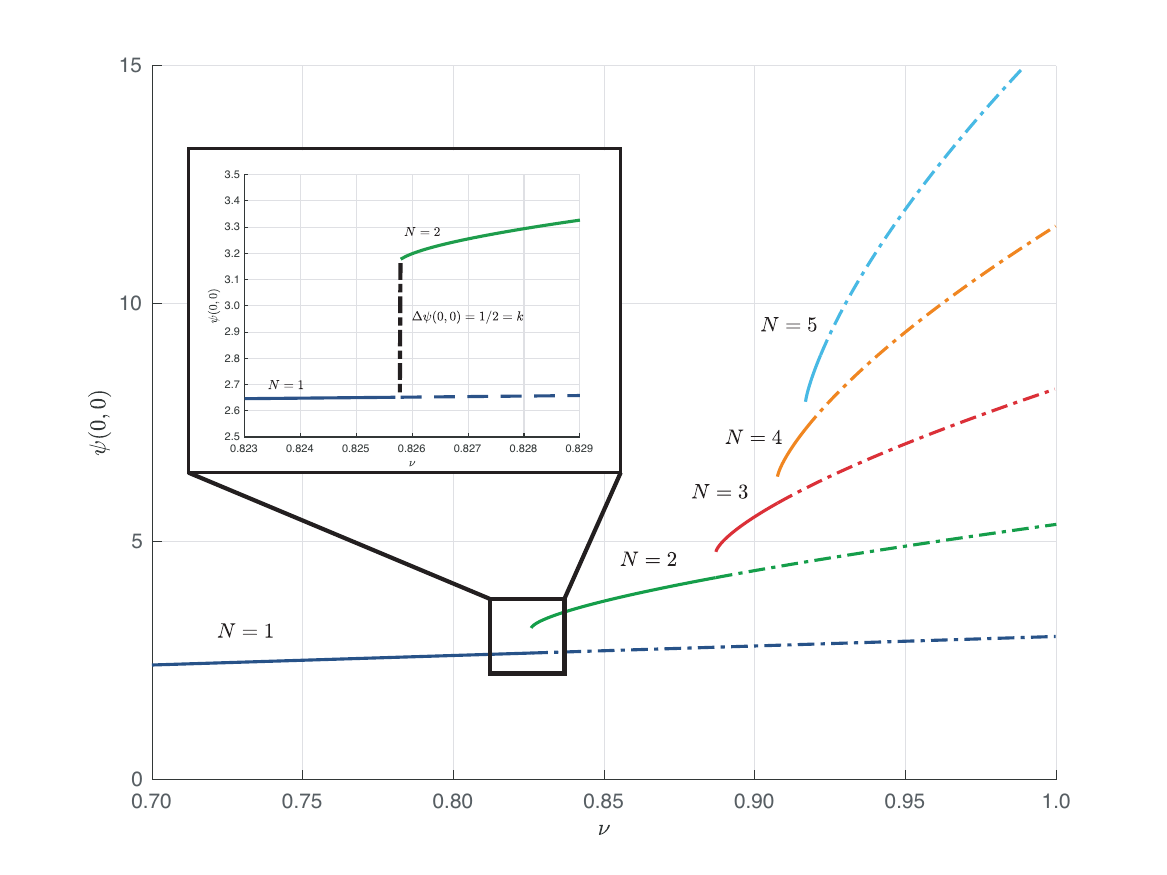}
	\caption{Plot of the peak height formula \eqref{eq:phfDn} at $k=1/2$ using the eigenvalues given by \eqref{eq:nu_m}. The insert shows a zoomed in version of the region between the peak heights of first and second order breathers, with a gap of $\Delta \psi(0,0)=1/2=k$. \label{fig:phfdn}}
\end{figure*}

Consequently, each curve $\nu_m$ in the $\nu\nu_m$-plane starts from the point:
\begin{align}
L_m(k) = \left(\nu_m^{*}(k), \frac{k}{2}\right) \quad , \quad m \geq 2	 \label{eq:lowerPoint}\,.
\end{align}
This is a stricter condition on $\nu$ than the lower-limit of Eq. \eqref{vran}, since $k > (1-\sqrt{1-k^2}) \, \forall \, k \in (0,1)$. Referring to the PHF \eqref{eq:phfDn}, the lower limit \eqref{eq:lowerPoint} then implies that when transitioning from an $N^\text{th}$-order to an $(N+1)^{\text{st}}$ order breather, the peak would instantly jump by $2(k/2) = k$ and would not just increase smoothly as in the simpler constant background case \cite{Chin2016}. This can be seen clearly in Fig. \ref{fig:phfdn}. 

\begin{figure*}[h] 
	\begin{minipage}{.5\linewidth}
		\centering
		\subfloat[]{\includegraphics[width=.95\linewidth]{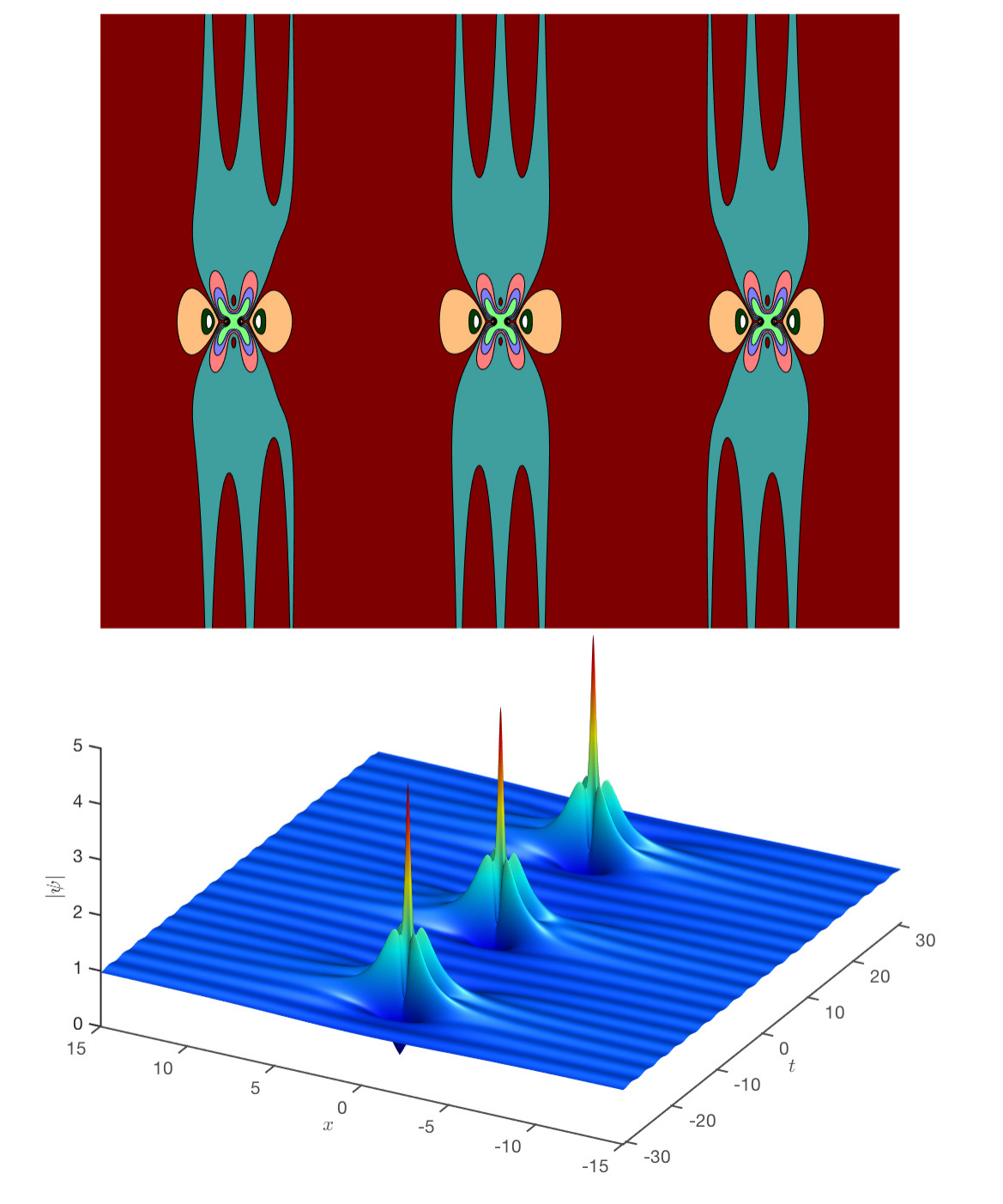}}
	\end{minipage}%
	\begin{minipage}{.5\linewidth}
		\centering
		\subfloat[]{\includegraphics[width=.95\linewidth]{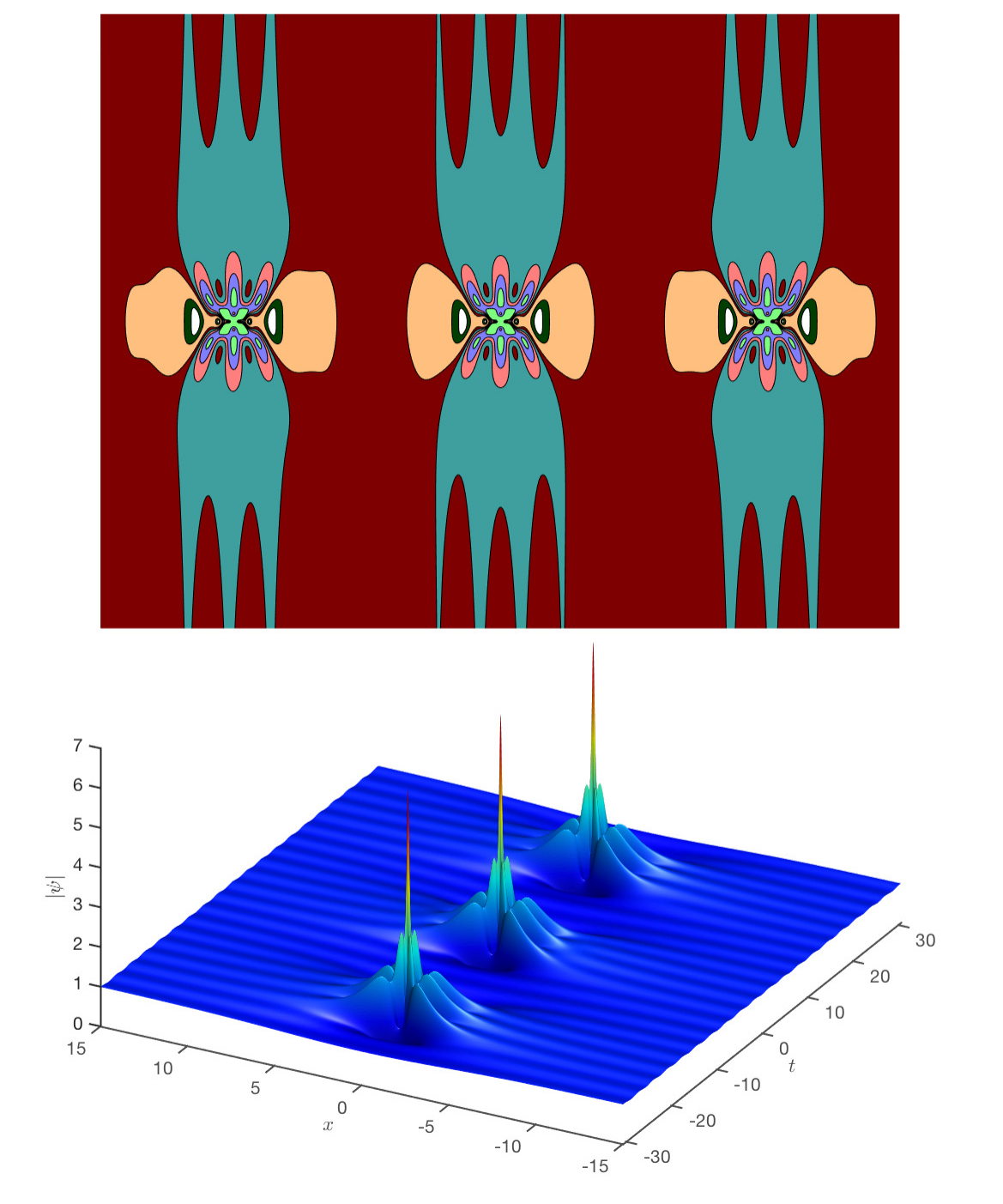}}
	\end{minipage}\par
	\begin{minipage}{.5\linewidth}
		\centering
		\subfloat[]{\includegraphics[width=.95\linewidth]{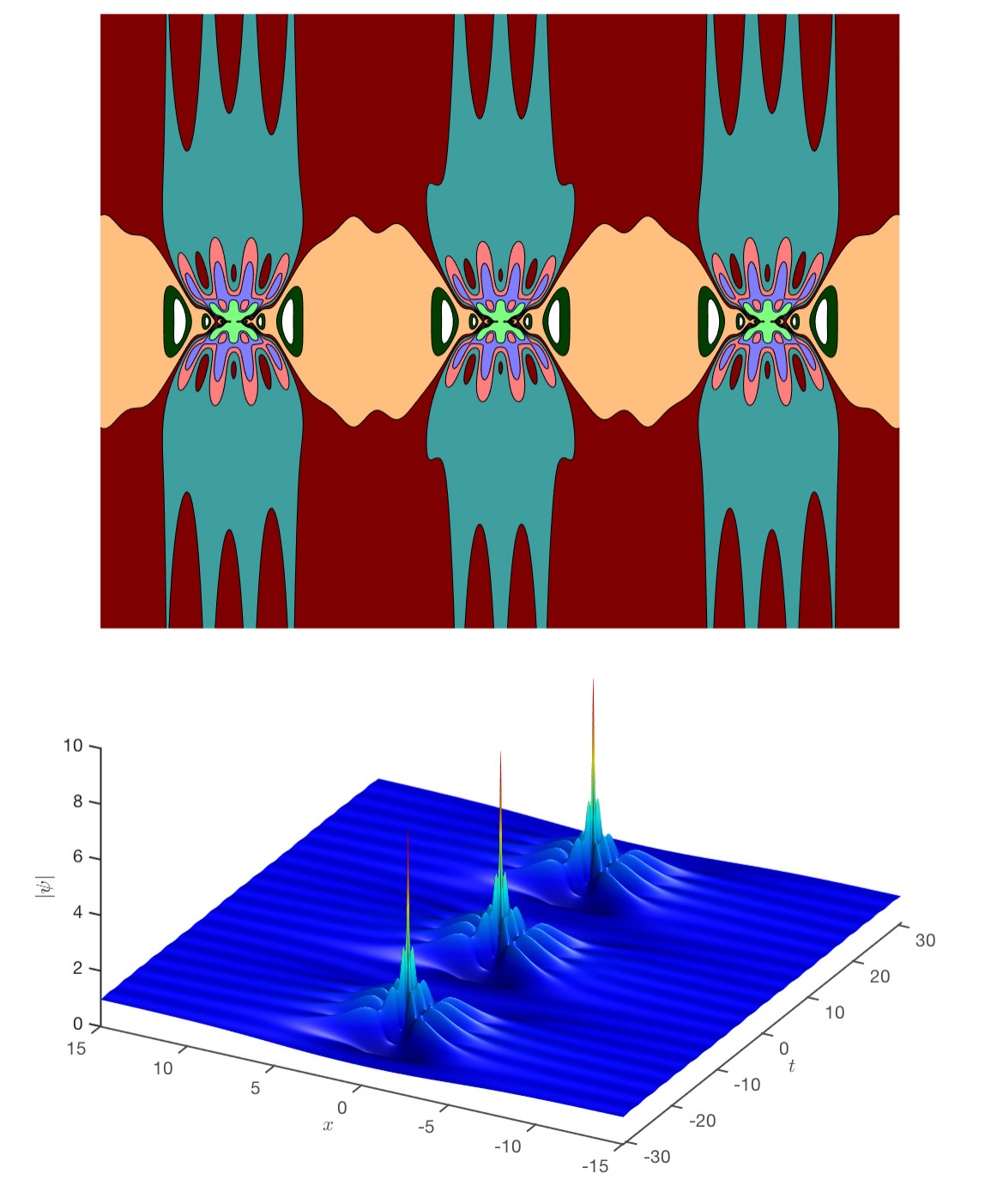}}
	\end{minipage}%
	\begin{minipage}{.5\linewidth}
		\centering
		\subfloat[]{\includegraphics[width=.95\linewidth]{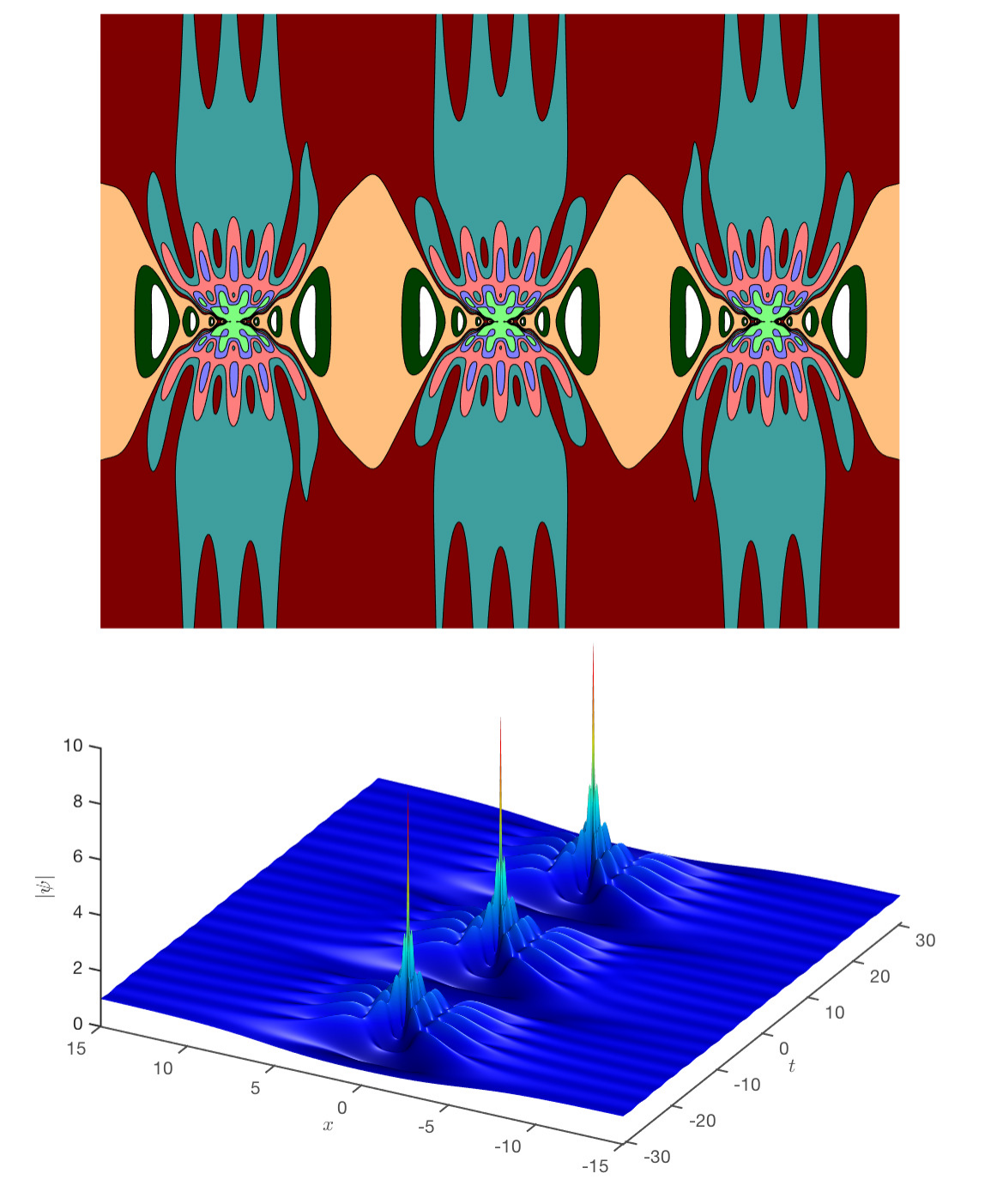}}
	\end{minipage}\par\medskip
	\caption{Quasi-periodic higher-order breathers with $(\nu \approx 0.972, k=1/4)$, and higher-order $\nu_m$ computed via Eq. \eqref{eq:nu_m} (a) Second order  (b) Third order (c) Fourth order (d) Fifth order. Inserts: contour plots to emphasize the low-$|\psi|$ details.  \label{fig:ABN_quasi}}
\end{figure*}

Figure \ref{fig:ABN_quasi} shows the three ``quasi-periods" of higher-order breathers with $\nu_m$ obtained from Eq. \eqref{eq:nu_m}, and $\nu$ and $k$ selected arbitrarily. The effect of matching the constituent breathers' periods drastically improves the overall periodicity of the high-order breather, with seemingly repeating higher-order side-peaks in the 3D plots. However, closer inspection of contour plots near the breather's base reveals remaining distortions in the side-peaks. They are still not identical to the central peak. This distortion is reminiscent of the first-order case, as shown in Fig. \ref{fig:AB1_unmatched}. These higher-order breathers are all quasi-periodic. 
In the limit of $k \rightarrow 0$, $\nu_m\rightarrow \sqrt{m^2-1}/m$,
(\ref{eq:nu_m}) reproduces our previous results of $a_m=\nu_m^2/2=(1-1/m^2)/2$ for breathers on a constant background in Ref. \cite{Chin2016}. 

\section{Fully Periodic Breathers\label{sec:fullyPeriodic}}

The period of the $\dn(t;k)$ background is given by:
\begin{align}
T_{\text{dn}} = 2K(k)
\label{eq:ellipK}
\end{align}
where $K(k) \equiv \int_0^\frac{\pi}{2} \frac{d\theta}{\sqrt{1-k^2 \sin^2\theta}}$ is the complete elliptic integral of the first kind. Matching the period of the background \eqref{eq:ellipK} to the breather requires:
\begin{align*}
T_{\text{B}} = q T_\text{dn}
\end{align*}
where $T_{\text{B}} = 2\pi/(2\kappa) = \pi/\kappa$ is the period of the fundamental breather and $q$ is a positive integer. This leads to an expression for $\kappa$:
\begin{align}
\kappa = \frac{\pi}{2 q K(k)}\,.
\label{eq:bgMatch}
\end{align}
Substituting back into \eqref{eq:kappa_dn}, one obtains the condition for matching a breather to the background: 
\begin{align}
\begin{aligned}
(\nu, k) \in \gamma_q := \Bigg\{&(\nu, k) \in (0,1)\times[0,1] \, \Bigg|\\ &\left(\frac{\pi}{2 q K(k)}\right)^2 + \frac{\left(k^2 + 4 \nu ^2\right)^2}{16 \nu^2} = 1\Bigg\}\,.
\end{aligned}
\label{eq:gammaContours}
\end{align}

\begin{figure*}[h]
	\centering
	\includegraphics[width=0.9\linewidth]{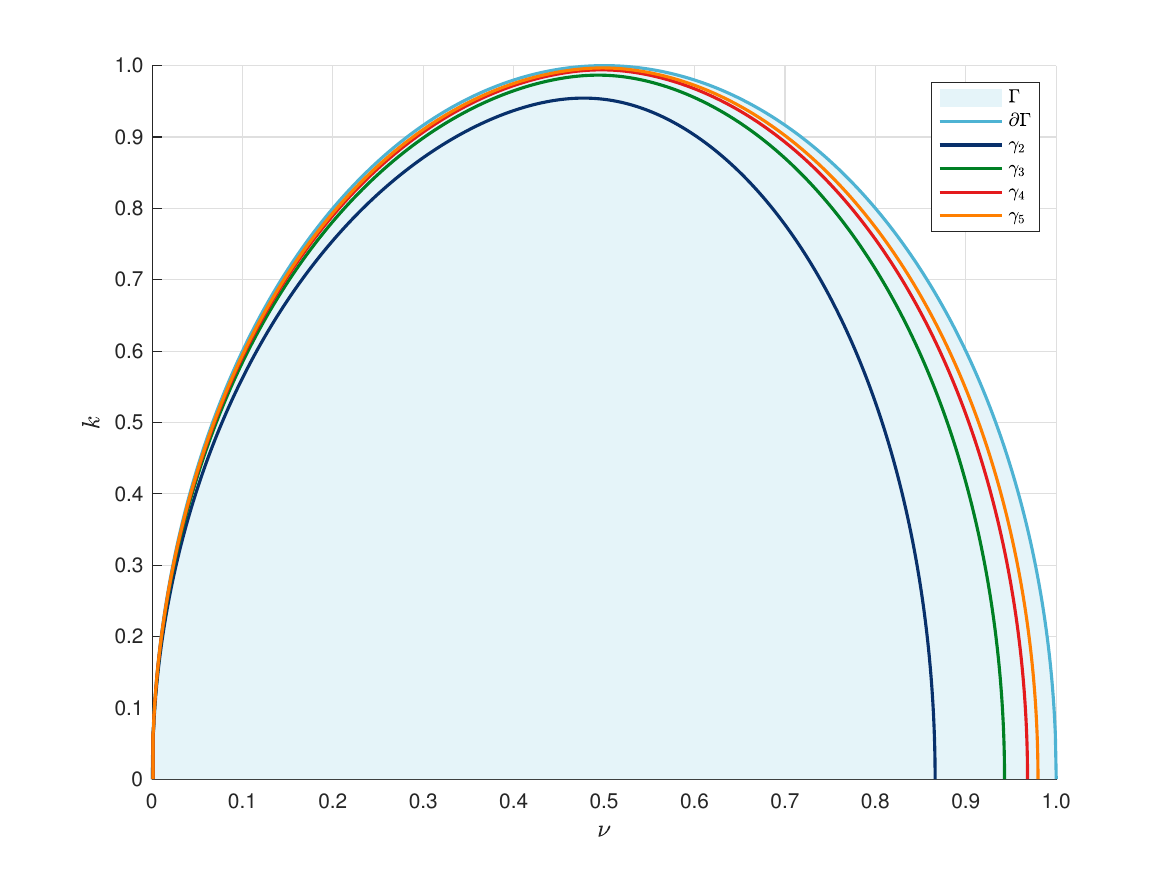}
	\caption{$\nu k$-plane, showing different regions of periodicity. The light-blue shaded region is where $\kappa$ is real, given by \eqref{eq:fundPeriod}. The solid curves correspond to different values of $q$ in \eqref{eq:gammaContours}. \label{fig:firstPeriodicity}}
\end{figure*}
Note that $\gamma_{q = 1} = \varnothing$ (i.e., one cannot match to the breather so that its period is exactly equal to the background, it must be a periodic multiple). As shown in Fig. \ref{fig:firstPeriodicity}, it is clear that $\gamma_q \subset \Gamma \, \forall \, q \in \mathbb{Z}^{+}$, and $\gamma_{q \rightarrow \infty} = \partial\Gamma$ (i.e. the boundary of $\Gamma$, Eq. \eqref{eq:fundPeriod}). Thus, any $(\nu,k) \in \gamma_q$ results in $\kappa \in \mathbb{R}$. Additionally, if a higher-order breather has matched constituents via Eq. \eqref{eq:nu_m}, its period will be $T_{B}$, and thus one only needs $(\nu,k) \in \gamma_q$. 

\begin{figure*}[h] 
	\centering
	\includegraphics[width=.95\linewidth]{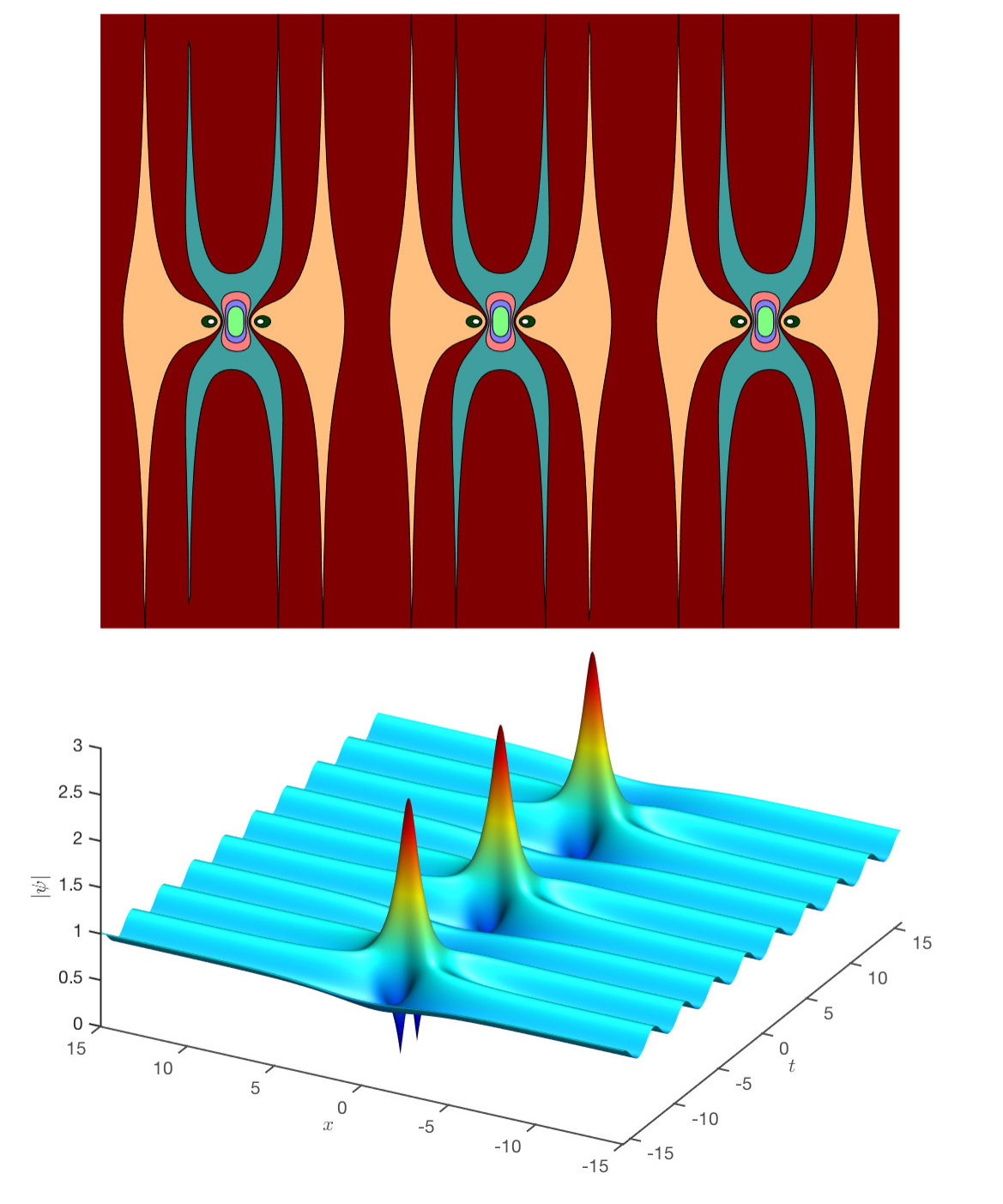}
	\caption{Fully periodic first-order breathers with $(\nu \approx 0.848, k = 3/5) \in \gamma_3$.  \label{fig:AB1_matched}}
\end{figure*}

\begin{figure*}[h] 
	\begin{minipage}{.5\linewidth}
		\centering
		\subfloat[]{\includegraphics[width=.95\linewidth]{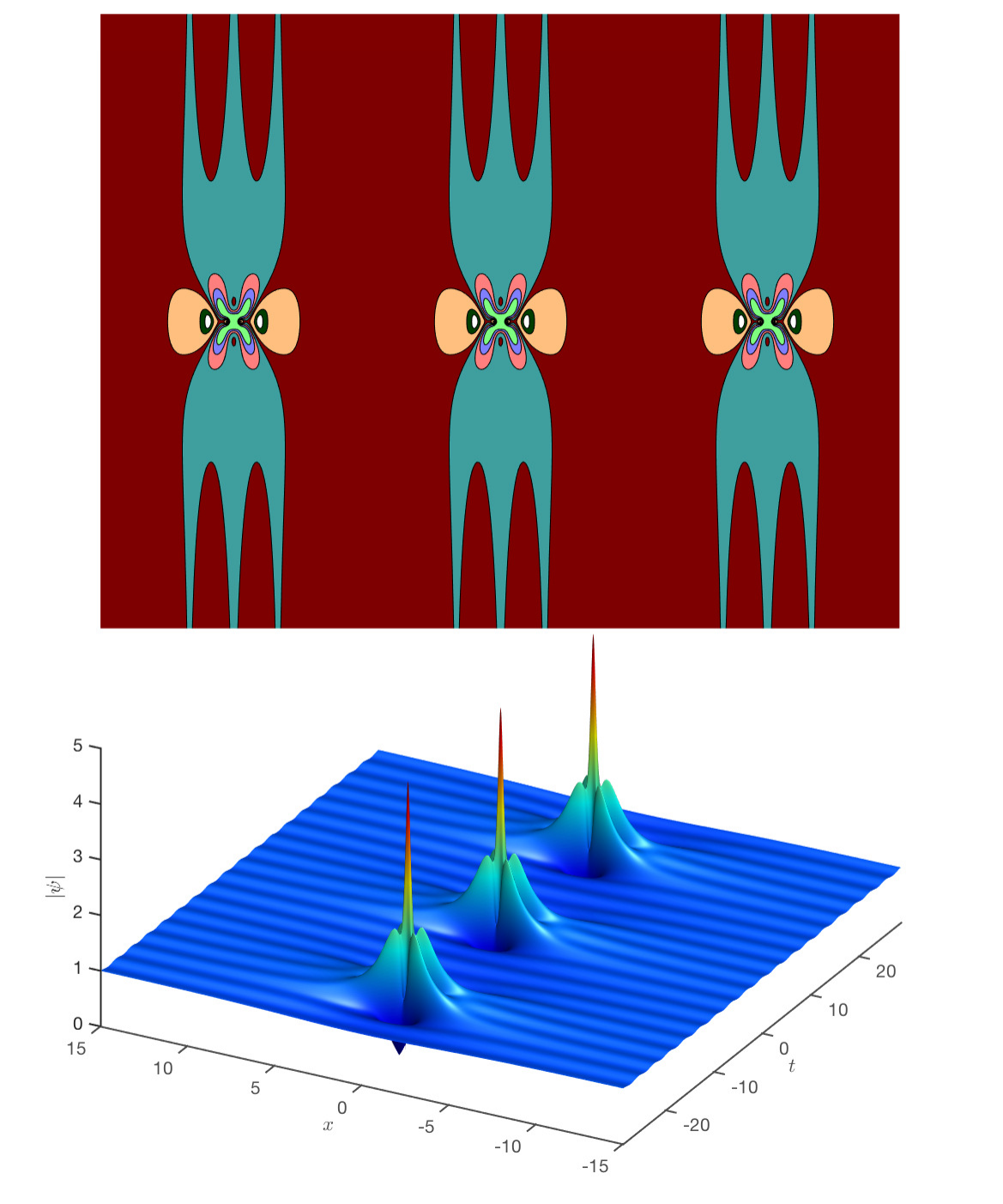}}
	\end{minipage}%
	\begin{minipage}{.5\linewidth}
		\centering
		\subfloat[]{\includegraphics[width=.95\linewidth]{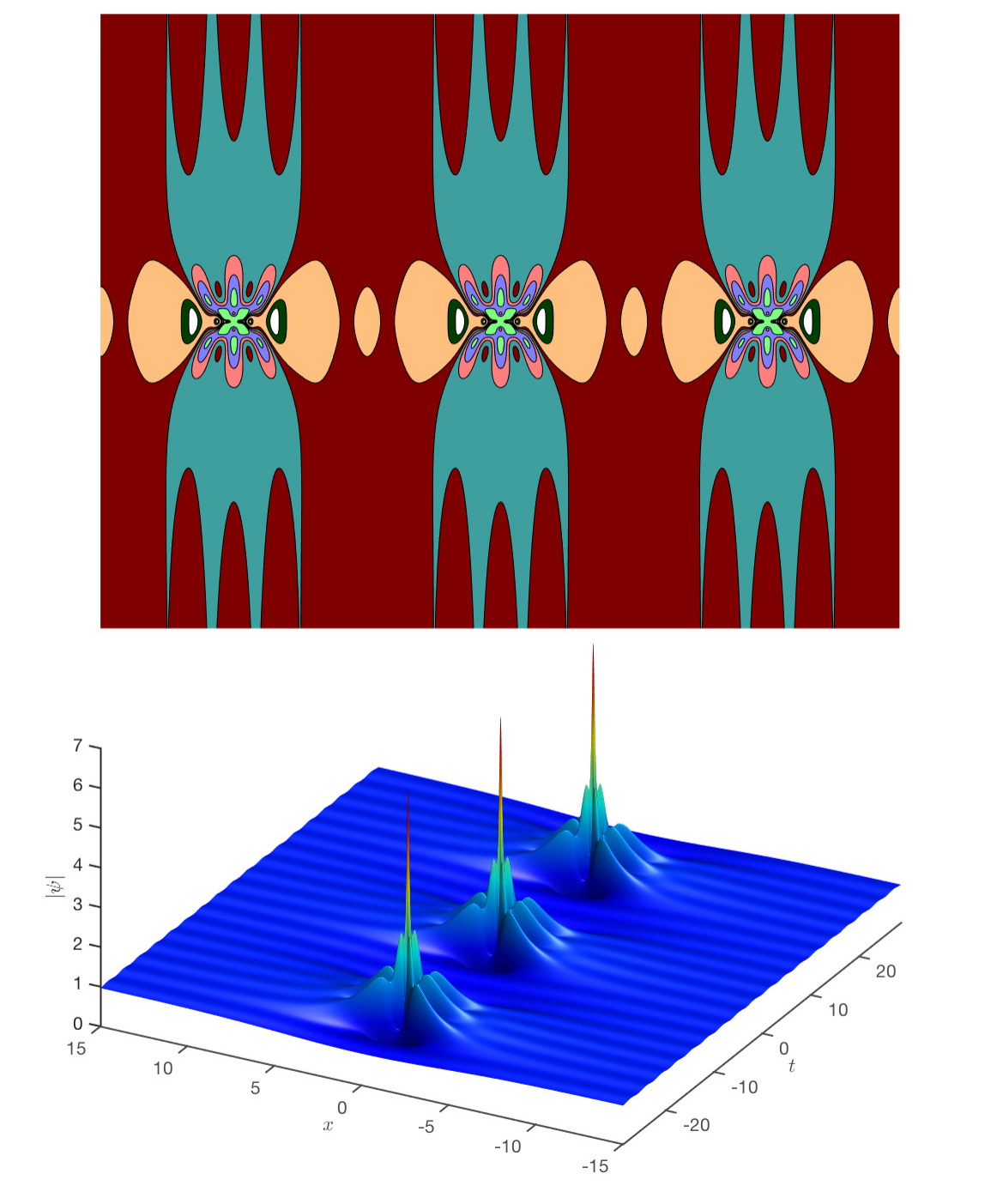}}
	\end{minipage}\par
	\begin{minipage}{.5\linewidth}
		\centering
		\subfloat[]{\includegraphics[width=.95\linewidth]{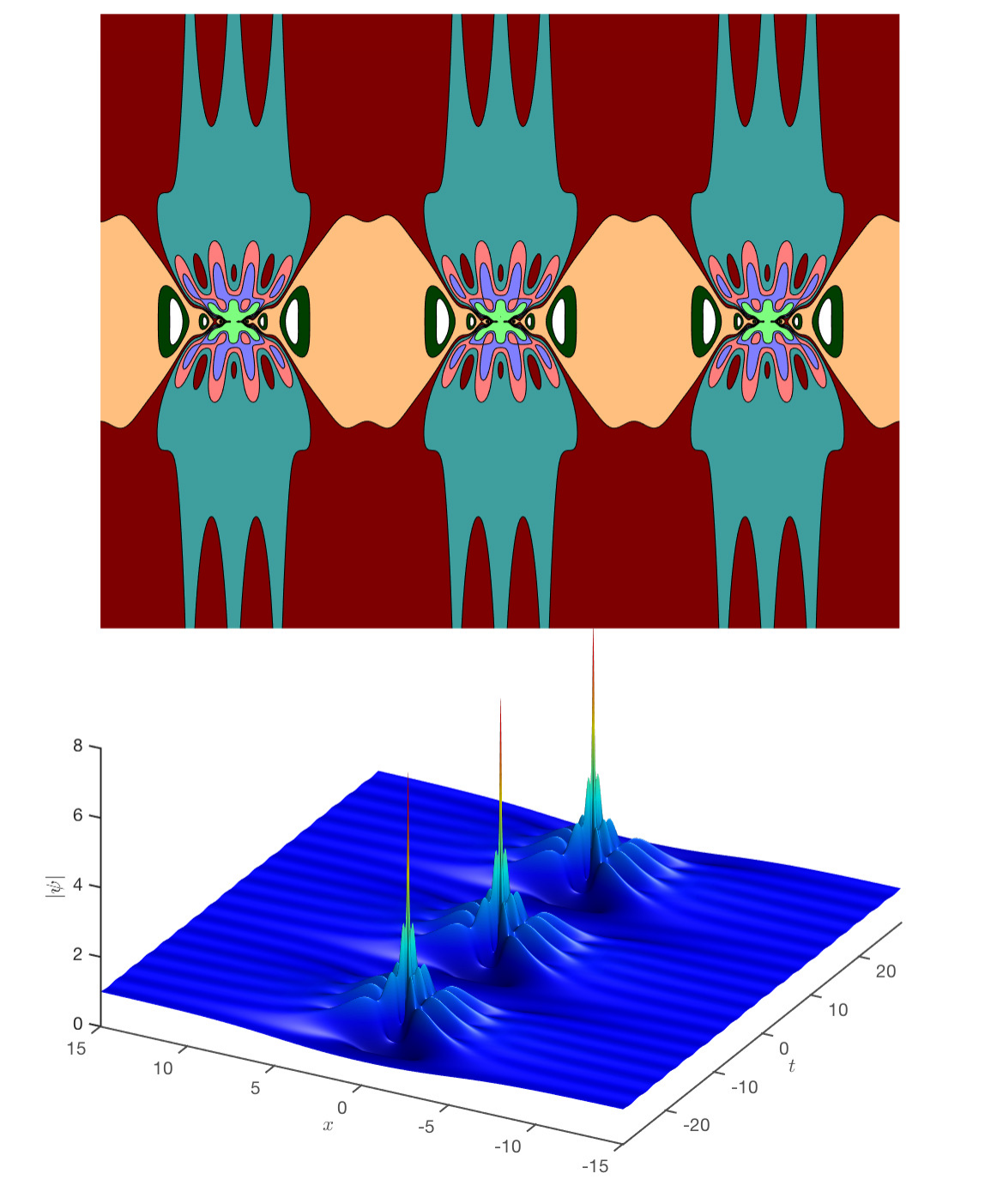}}
	\end{minipage}%
	\begin{minipage}{.5\linewidth}
		\centering
		\subfloat[]{\includegraphics[width=.95\linewidth]{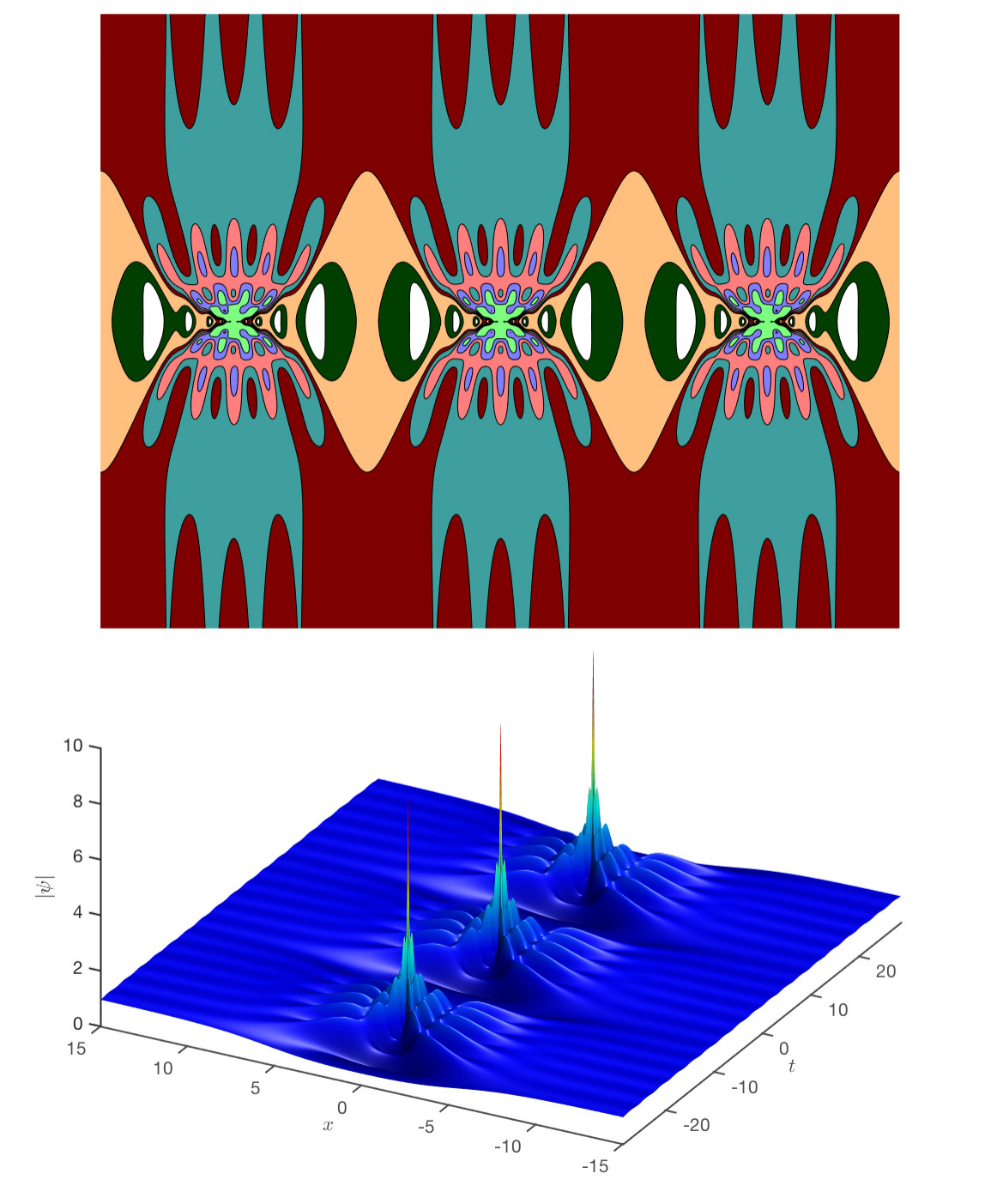}}
	\end{minipage}\par\medskip
	\caption{Fully periodic higher-order breather with $(\nu \approx 0.9704, k = 1/4) \in \gamma_6$, and the higher $\nu_m$ calculated via Eq. \eqref{eq:nu_m}. (a) Second order. (b) Third order (c) Fourth order. (d) Fifth order. Insert: Contour plots to emphasize the low-$|\psi|$ detail. The symmetry between the central and side peaks is apparent. \label{fig:ABN_matched}}
\end{figure*}

Figure \ref{fig:AB1_matched} shows a fully periodic first-order breather. It is clear that selecting a set of parameters $(\nu, k)$ satisfying Eq. \eqref{eq:gammaContours} leads to a truly periodic structure. We additionally show several fully periodic higher-order breathers in Fig. \ref{fig:ABN_matched}, combining conditions \eqref{eq:gammaContours} and \eqref{eq:nu_m}. In general, the effect of matching to the background is subtle in 3D plots yet abundantly clear in the contour plots. The central breather's peak is centered at the origin, where $\psi_0(0,0) = 1$. If the parameters $(\nu,k) \notin \gamma_q$ (i.e. $q \notin \mathbb{Z}$), the side peaks will then be displaced from the peak of $\psi_0$. On the other hand, when $(\nu,k) \in \gamma_q, \, q \in \mathbb{Z}^{+}$, all breather peaks lie precisely on top of the background crests, and we get perfect periodicity. We provide a video in this work's supplemental material, which visually demonstrates this process.

\section{Discussion and Conclusion \label{sec:conclusion}}

In this work, we have numerically solved for the higher-order breathers of the nonlinear Schr\"odinger equation on an elliptic background by iterating the Darboux transformation. We find that the periodic background exerts a much more profound impact on the higher-order breathers than previously thought. Because of the periodic background, any unmatched higher-order breather has only a single high-intensity peak, resembling a rogue wave. 
These QRWs have a far richer structure than the standard, monotonic Peregrine-like rational rogue waves. 

The chaotic background shown in Fig. \ref{fig:ABN_unmatched} is also of interest. It is well known that two-wave mixing can only result in quasi-periodic structures.
This is the case of Fig. \ref{fig:AB1_unmatched}, the two waves being waves having the breather's period and that of the background.
However, three and more wave mixing can lead to chaos. This is the resulting chaotic-looking background we see in Fig. \ref{fig:ABN_unmatched}. 

Truly periodic higher-order breathers can only be recovered by matching all the constituent breathers' periods to each other and the background. Matching constituent breathers' periods to each other produces quasi-periodic with distorted side peaks. Further matching the breathers' periods to that of the background yields side-peaks identical to the central peak, sitting precisely on top of the background's crests some periodic distance away. While a general breather of order $N$ has an $N+1$ dimensional parameter space, fully periodic breathers are constrained to a countably-infinite set of contours in a 2D parameter space, thus being exceedingly rare. We have termed such a periodic higher-order breather the periodic rogue wave in Ref. \cite{Nikolic2018}. Our findings are summarized in Table \ref{tab:summary}. This explains why, on a periodic background, the only common high peak structures that can occur are, paradoxically, the quasi-rogue waves of Fig. \ref{fig:ABN_unmatched}. 

\begin{table*}
\begin{center}
\caption{\label{tab:summary} Necessary and sufficient conditions on the parameters for a fully periodic breather of order $N$. }
 \begin{tabular}{c c c} 
 \hline
 Condition & Equation & Explanation \\ [0.5ex] 
 \hline
 $(\nu, k) \in \gamma_{q}, \,  q = 2, 3, \ldots $ & Eq. \eqref{eq:gammaContours} & Matches a breather to the background\\
  $\nu \equiv \nu_1 > \nu_N^*$ & Eq. \eqref{eq:nudn_star} & Ensures $\nu_m \in \mathbb{R}, m = 1, \ldots, N$\\
 $\nu_m = \nu_m(\nu, k), m = 1, \ldots, N$ & Eq. \eqref{eq:nu_m} & Matches the constituent breathers to each other \\
 \hline
\end{tabular}
\end{center}
\end{table*}

\section*{Supplementary Material}
See supplementary material for a video demonstrating the matching of the breathers to the dnoidal background.

This research is supported by the Qatar National Research Fund (Project NPRP 8-028-1-001). O.A.A. is supported by the Berkeley Graduate Fellowship and the Anselmo J. Macchi Graduate Fellowship. S.N.N. acknowledges support from Grants III 45016 and OI 171038 of the Serbian Ministry of Education, Science and Technological Development. M.R.B. acknowledges support by the Al-Sraiya Holding Group.

\appendix
\section{The Darboux transformation and the Peak Height Formula \label{sec:phf}}

The cubic NLSE \eqref{eq:nlse} can be written as the compatibility condition of the following two equations \cite{Akhmediev1997}:

\begin{align}
\begin{aligned}
R_t &= \bs{L}R, \quad R_x &= \bs{A}R,
\end{aligned}
\label{eq:lax}
\end{align}
where
\begin{align}
R &= 
\begin{pmatrix}
r\\
s
\end{pmatrix} \equiv
\begin{pmatrix}
r_{1m}\\
s_{1m}
\end{pmatrix} \, , \\ 
\bs{L} &= 
\begin{pmatrix}
-i\lambda & \psi \\
-\psi^* & i\lambda \\
\end{pmatrix} \, , \,
\bs{A} = 
\begin{pmatrix}
-i\lambda^2+i\half|\psi|^2 & \lambda\psi + i\half \psi_t \\
-\lambda\psi^* + i\half \psi_t^* & i\lambda^2-i\half|\psi|^2\\
\end{pmatrix}.
\end{align}
The operators (matrices) $\bs{L}$ and $\bs{A}$ are known as the Lax pair of \eqref{eq:nlse}, and the functions $r(x,t)$ and $s(x,t)$ are the Lax pair generating functions. $\lambda$ is generally a complex eigenvalue, and is independent of the evolution variable $x$ (so that the Lax pair is isospectral). The compatibility condition of \eqref{eq:lax} is known as the Lax equation or zero-curvature condition, and gives rise to the cubic NLSE \eqref{eq:nlse}:
\begin{align}
\bs{L}_x-\bs{A}_t - [\bs{A},\bs{L}] = \bs{0} \, ,
\label{eq:zeroCurv}
\end{align}
where $[\bs{A},\bs{L}] \equiv \bs{AL}-\bs{LA}$ is the commutator. 

Given an initial solution of \eqref{eq:nlse}, one can find a more complicated solution of order $N$ via the Darboux transformation:  \cite{Akhmediev2009c,Kedziora2011, Matveev1991, Akhmediev1997}:

\begin{align}
\psi_N = \psi_0 + \sum_{m=1}^{N}\frac{2 i r_{m1} s_{m1}^* \left(\lambda_m -\lambda_m ^*\right)}{|r_{m1}|^2+|s_{m1}|^2}\,,
\label{eq:DTIter2}
\end{align}
where the higher-order Lax-pair generating functions are computed recursively via \cite{Akhmediev1997,Matveev1991}:

\begin{align}
r_{mj}&=[(l_{m-1}^*-l_{m-1})s^*_{m-1,1}r_{m-1,1}s_{m-1,j+1}\nonumber\\
&+(l_{j+m-1}-l_{m-1})|r_{m-1,1}|^2r_{m-1,j+1}\nonumber\\
 &+(l_{j+m-1}-l^*_{m-1})|s_{m-1,1}|^2r_{m-1,j+1}] \nonumber \\
 &/(|r_{m-1,1}|^2+|s_{m-1,1}|^2),\label{eq:r_rec}\\
s_{mj}&=[  (l_{m-1}^*-l_{m-1})s_{m-1,1}r^*_{m-1,1}r_{m-1,j+1}\nonumber\\
&+(l_{j+m-1}-l_{m-1})|s_{m-1,1}|^2s_{m-1,j+1}\nonumber\\
&+(l_{j+m-1}-l^*_{m-1})|r_{m-1,1}|^2s_{m-1,j+1}] \nonumber \\
&/(|r_{m-1,1}|^2+|s_{m-1,1}|^2).
\label{eq:s_rec}
\end{align}
To find the starting functions of the recursion, i.e. $r_{1m}$ and $s_{1m}$, we take the Ans\"atze:
\begin{align}
\begin{aligned}
r_{1m}(x,t) &= a_{1m}(x,t)e^{\frac{ix}{4}\left(k^2-2\right)}\,,\\
s_{1m}(x,t) &= a_{1m}(x,t)e^{-\frac{ix}{4}\left(k^2-2\right)}\,,
\end{aligned}
\end{align}
and $\psi$ as given in Eq. \eqref{eq:dnseed}. Substituting in the Lax pair equation \eqref{eq:lax} and suppressing subscripts, one gets \cite{Kedziora2014}:
\begin{align}
\begin{aligned}
a_t &= i\lambda a(x,t) + ib(x,t) \dn(t;k), \\
b_t &= -i\lambda b(x,t) + ia(x,t)\dn(t;k),\\
a_x &= \half i a(x,t) \left(2\lambda^2 + k^2\left(\sn^2(t;k) - \half\right)\right)\\
    &+ b(x,t)\left(i\lambda\dn(t;k) - \frac{k^2}{2}\sn(t;k)\cn(t;k)\right), \\
b_x &= -\half i b(x,t) \left(2\lambda^2 + k^2\left(\sn^2(t;k) - \half\right)\right)\\
    &+ a(x,t)\left(i\lambda\dn(t;k) + \frac{k^2}{2}\sn(t;k)\cn(t;k)\right).
\end{aligned}
\label{eq:abdn}
\end{align}
However, as noted in \cite{Kedziora2014}, the four coupled first-order differential equations \eqref{eq:abdn} cannot be solved analytically. Nonetheless, one can solve for the profiles and derivatives at $t = 0$, given by \cite{Kedziora2014}:
\begin{align}
\begin{aligned}
a_{1m}|_{t=0} &= Ae^{i(\chi_m+\kappa_m\lambda_m(x-x_m))} - Be^{-i(\chi_m+\kappa_m\lambda_m(x-x_m))},\\
b_{1m}|_{t=0} &= A e^{i(-\chi_m+\kappa_m\lambda_m(x-x_m))} + Be^{-i(-\chi_m+\kappa_m\lambda_m(x-x_m))},\\
a_{1m,t}|_{t=0} &= i(\lambda_m a_{1m}|_{t=0} + b_{1m}|_{t=0}),\\
b_{1m,t}|_{t=0} &= -i(\lambda_m b_{1m}|_{t=0} - a_{1m}|_{t=0}),
\end{aligned}
\label{eq:abprofiles}
\end{align}
where 
\begin{align}
\kappa _m=\sqrt{\left(\lambda _m-\frac{k^2}{4 \lambda _m}\right)^2+1}
\end{align}
is half the wavenumber of the \mth constituent breather, $\chi_m = \arccos{(\kappa_m)}/2$ is the background dependent phase, and $A$ and $B$ are two phase constants. One can then evolve Eq. \eqref{eq:abprofiles} along the $t$-axis numerically. All the results in this work employ a fourth-order Runge-Kutta method and sufficiently small grid spacing.

In previous works \cite{Chin2016,Chin2016a,Nikolic2017}, using the DT, we derived the so-called Peak Height Formula (PHF) for $N^{\text{th}}$-order solutions of nonlinear Schr\"odinger-type equations:

\begin{equation}
\psi_N(0,0) = \psi_0(0,0) +2 \sum_{m=1}^{N} \nu_m
\label{eq:phf}
\end{equation}
This requires the Lax pair generating functions to only differ by a phase at the origin \cite{Chin2016a}:
\begin{align}
s_{1m}(0,0) = e^{i \phi}r_{1m}(0,0),
\label{eq:phfCondition}
\end{align}
One can easily verify that equations \eqref{eq:abprofiles}, with proper choice of $A$ and $B$, lead to $r(x,t)$ and $s(x,t)$ that obey the PHF condition \eqref{eq:phfCondition} with $\phi$ = $\pi/2$, leading to the PHF \eqref{eq:phfDn}.

\bibliographystyle{spmpsci}      

\bibliography{main}
\end{document}